\documentclass{aa}

\usepackage{graphicx}
\usepackage{txfonts}
\usepackage{natbib}
\usepackage{amsmath}
\usepackage{booktabs}
\usepackage{array}
\usepackage{siunitx}
\usepackage[colorlinks=true,allcolors=blue]{hyperref}

\newcommand{\ffdeg}{\mbox{\ensuremath{.\!\!\degr}}}

\newcommand{\ffarcs}{\mbox{\ensuremath{.\!\!^{\prime\prime}}}}

\newcolumntype{L}[1]{>{\raggedright\let\newline\\\arraybackslash\hspace{0pt}}m{#1}}
\newcolumntype{C}[1]{>{\centering\let\newline\\\arraybackslash\hspace{0pt}}m{#1}}

\bibpunct{(}{)}{;}{a}{}{,}

\begin{document}

\title{Scattered light mapping of protoplanetary disks}

\author{
T.~Stolker\inst{1}
\and C.~Dominik\inst{1}
\and M.~Min\inst{2,1}
\and A.~Garufi\inst{3,4}
\and G.~D.~Mulders\inst{5,6}
\and H.~Avenhaus\inst{7,8}
}

\institute{
Anton Pannekoek Institute for Astronomy, University of Amsterdam, Science Park 904, 1098 XH Amsterdam, The Netherlands\\
\email{T.Stolker@uva.nl}
\and SRON Netherlands Institute for Space Research, Sorbonnelaan 2, 3584 CA Utrecht, The Netherlands
\and Universidad Aut\'{o}noma de Madrid, Dpto. F\'{i}sica Te\'{o}rica, M\'{o}dulo 15, Facultad de Ciencias, Campus de Cantoblanco, E-28049 Madrid, Spain
\and Institute for Astronomy, ETH Zurich, Wolfgang-Pauli-Strasse 27, 8093 Zurich, Switzerland
\and Lunar and Planetary Laboratory, The University of Arizona, Tucson, AZ 85721, USA
\and Earths in Other Solar Systems Team, NASA Nexus for Exoplanet System Science
\and Departamento de Astronom\'{i}a, Universidad de Chile, Casilla 36-D, Santiago, Chile
\and Millennium Nucleus "Protoplanetary Disks", Chile
}

\date{Received ?; accepted ?}

\abstract {High-contrast scattered light observations have revealed the surface morphology of several dozens of protoplanetary disks at optical and near-infrared wavelengths. Inclined disks offer the opportunity to measure part of the phase function of the dust grains that reside in the disk surface which is essential for our understanding of protoplanetary dust properties and the early stages of planet formation.}
{We aim to construct a method which takes into account how the flaring shape of the scattering surface of an (optically thick) protoplanetary disk projects onto the image plane of the observer. This allows us to map physical quantities (scattering radius and scattering angle) onto scattered light images and retrieve stellar irradiation corrected ($r^2$-scaled) images and dust phase functions.}
{The scattered light mapping method projects a power law shaped disk surface onto the detector plane after which the observed scattered light image is interpolated backward onto the disk surface. We apply the method on archival polarized intensity images of the protoplanetary disk around HD~100546 that were obtained with VLT/SPHERE in $R'$-band and VLT/NACO in $H$- and $K_{\rm s}$-band.}
{The brightest side of the $r^2$-scaled $R'$-band polarized intensity image of HD~100546 changes from the far to the near side of the disk when a flaring instead of a geometrically flat disk surface is used for the $r^2$-scaling. The decrease in polarized surface brightness in the scattering angle range of $\sim$40\degr\,--\,$70\degr$ is likely a result of the dust phase function and degree of polarization which peak in different scattering angle regimes. The derived phase functions show part of a forward scattering peak which indicates that large, aggregate dust grains dominate the scattering opacity in the disk surface.}
{Projection effects of a protoplanetary disk surface need to be taken into account to correctly interpret scattered light images. Applying the correct scaling for the correction of stellar irradiation is crucial for the interpretation of the images and the derivation of the dust properties in the disk surface layer.}


\maketitle

\section{Introduction}\label{sec:introduction}

High-contrast scattered light observations of protoplanetary disks have revealed intriguing morphologies such as spiral arms, gaps, asymmetries, and shadows \citep[e.g.,][]{wisniewski2008,mayama2012,quanz2013,grady2013,wagner2015,rapson2015,stolker2016}, which may be signposts for disk evolution and planet-disk interactions \citep[e.g.,][]{pinilla2015,zhu2015,dong2015,rosotti2016}. Inclined disks offer the opportunity to measure the dust scattering efficiency at different angles with the star (i.e., the phase function) which is essential for our understanding of the properties and evolution of the dust in the disk surface. Projection effects are important for inclined disk surfaces, however, it is common practice to use a geometrically flat disk for the calculation of stellar irradiation corrected ($r^2$-scaled) images and phase functions \citep[e.g.,][]{quanz2011,garufi2014,thalmann2015}.

Planet formation is a complex process which requires sub-micron sized dust grains from the interstellar medium to grow 14 orders of magnitude in size towards planets \citep[e.g.,][]{armitage2013}. Dust grains will coagulate, settle, drift, and fragment depending on many aspects of the disk structure and dust properties \citep{testi2014}. Protoplanetary disks are optically thick at optical and near-infrared wavelengths, consequently, scattered light observations probe the dust in the surface layer. In the disk surface, dust grains are expected to be (sub-)micron sized, because large compact grains will settle and grow efficiently towards the midplane leading to a vertical stratification of dust grain sizes \citep{dubrulle1995,dullemond2004}. However, there are indications that also larger dust grains can be present in the disk surface \citep{mulders2013,stolker2016} which have presumably an aggregate structure that provides them with aerodynamic support against settling.

Polarimetric differential imaging (PDI) is a powerful technique to image protoplanetary and debris disks in scattered light at high angular resolution. The unpolarized speckle halo is removed with a differential linear polarization measurement which allows for high-contrast observations of disks that are multiple orders of magnitude fainter in scattered light compared to the stellar light. The scattered light surface brightness distribution of a disk depends on disk properties such as the pressure scale height and surface density, as well as the single scattering albedo, phase function, and single scattering polarization of the dust grains.

Interpretation of scattered light images of inclined protoplanetary disks can be non-trivial for several reasons. Firstly, the surface layer of a protoplanetary disks has usually a flaring shape which results in complex projection effects. Secondly, the surface brightness is partially scattering angle dependent because of the phase function and single scattering polarization of the dust grains. Thirdly, the stellar irradiation of the disk surface scales with the reciprocal of the squared distance. All these effects have to be taken into account for a correct interpretation of scattered light images and phase functions.

In this work, we will investigate the effect of the scattering surface geometry of a protoplanetary disk on the calculation of \mbox{$r^2$-scaled} images and phase functions. In Sect.~\ref{sec:mapping}, we construct a numerical method which corrects scattered light images of inclined and flaring disks for the dilution of the stellar radiation field and retrieves the phase function of the dust. In Sect.~\ref{sec:HD100546}, we apply the method on polarized intensity images of the HD~100546 protoplanetary disk. In Sect.~\ref{sec:discussion}, we discuss the importance of the method for the interpretation of polarized scattered light observations of HD~100546 and the dust grain properties in the disk surface in particular. In Sect.~\ref{sec:conclusions}, we summarize the main conclusions.

\section{Mapping of scattered light images}\label{sec:mapping}

\subsection{Photon scattering}\label{sec:photon_scattering}

The orientation of a protoplanetary disk surface with respect to the observer determines the local scattering angle of the stellar photons that irradiate the dust grains. The scattering geometry depends on the inclination of the disk and the vertical extent of the disk surface which both have to be considered for a correct interpretation of scattered light images. The optically thick part of the disk surface that is probed in scattered light is equal to the disk height where the scattering optical depth in radial direction from the star reaches unity ($\tau=1$). In general, this will be several times the local pressure scale height and will often be determined by the height of the inner edge of the disk which can be puffed up by stellar heating \citep{dullemond2001}.

Changes in surface brightness of an inclined disk are often related to the scattering angle at which incoming photons scatter from dust grains towards the observer. The scattering angle, $\Psi$, for a disk surface with constant opening angle, $\gamma$, is given by \citep{quanz2011}
\begin{equation}
\begin{aligned}
\Psi &= 90\degr + (i-\gamma) \cos{\phi} \qquad (-90\degr \leq \phi \leq 90\degr), \\
\Psi &= 90\degr + (i+\gamma) \cos{\phi} \qquad (90\degr > \phi < 270\degr),
\end{aligned}
\end{equation}
where $i$ is the disk inclination, $\gamma$ the disk opening angle, and $\Phi$ the azimuthal coordinate in the image plane measured in counterclockwise direction away from the far side minor axis of the disk. The phase function of the disk can be obtained by calculating the mean flux as function of scattering angle for all pixels within a certain distance range from the star in order to exclude any irradiation effect. Therefore, it is important to know the scattering radius and scattering angle associated with each image pixel. The scattering radius is the distance between the star and the scattering location in the disk surface and the scattering angle is the angle between the photon direction before and after scattering.

To first order, the radial distance from each pixel to the star can be obtained by correcting for the inclination of the disk only, which would correspond to the true distance in case the disk would be geometrically flat. For a more precise analysis, the flaring shape of the radial $\tau=1$ surface has to be considered which can have a significant effect on the radial distance. More specifically, the observer's view on the near and far side of an inclined and flaring disk will be grazing and frontal, respectively, which will increase and decrease, respectively, the calculated distance compared to the value which only takes into account the disk inclination.

\subsection{Flaring disk projection}\label{sec:disk_projection}

\begin{figure}
\centering
\resizebox{\hsize}{!}{\includegraphics{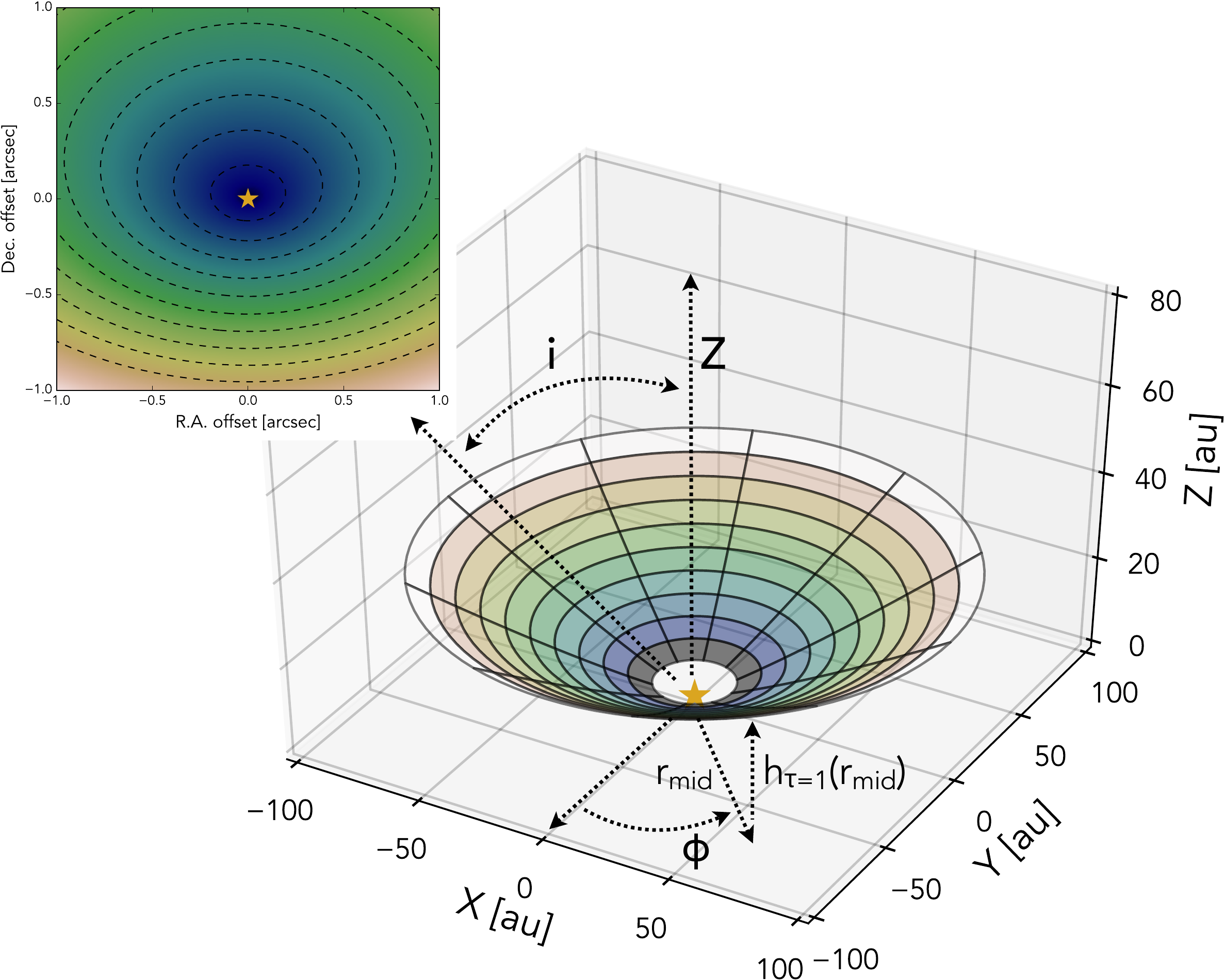}}
\caption{Schematic of the scattered light mapping method. A power law shaped disk surface is projected onto an image plane which has an inclination, $i$, and is located along the negative $y$-axis. The height of the radial $\tau=1$ surface, $h_{\rm \tau=1}(r_{\rm mid})$, at a midplane radius, $r_{\rm mid}$, is given by Eq.~\ref{eq:powerlaw}. The figure shows a disk projection with $i=42\degr$, ${\rm PA}=90\degr$, $h_0=0.14$, $\chi=1.17$, and a disk inner radius of 15~au at a distance of 100~pc.}
\label{fig:schematic}
\end{figure}

\begin{figure*}
\centering
\includegraphics[width=\textwidth]{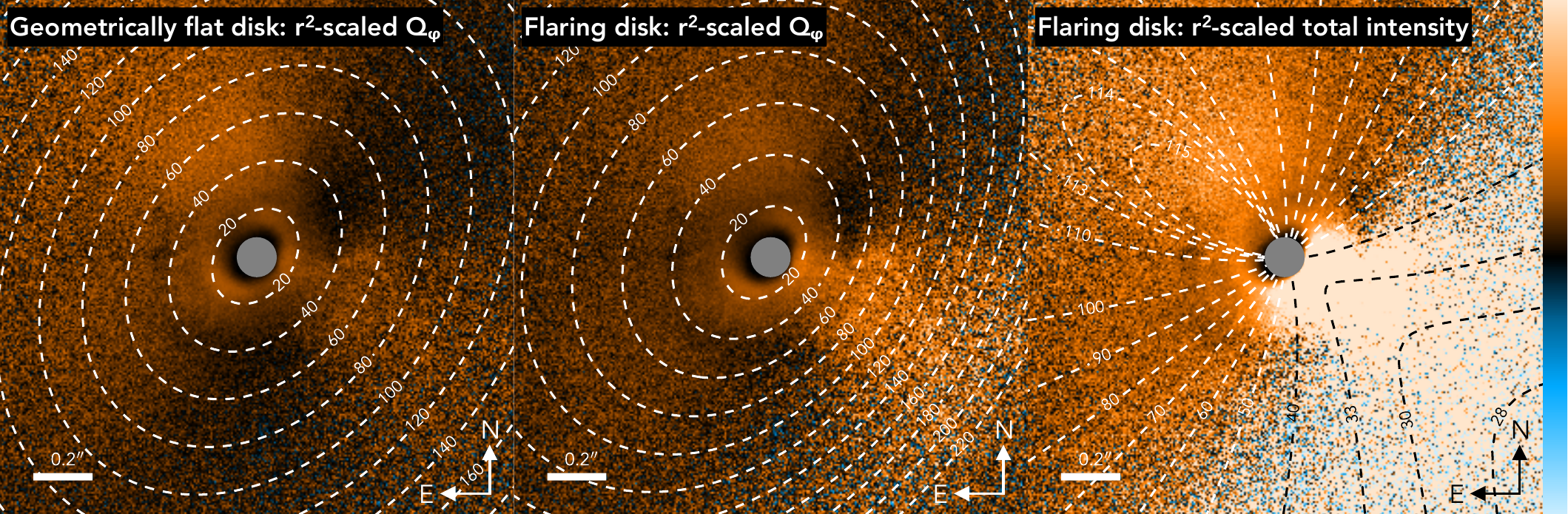}
\caption{\textbf{Left:} VLT/SPHERE coronagraphic $R'$-band $Q_\phi$ image of HD~100546 \citep{garufi2016}. The image is \mbox{$r^2$-scaled} with only a correction for the inclination of the disk. \textbf{Center}: The same $Q_\phi$ image with a correction for both the inclination and height of the flaring disk surface. \textbf{Right:} Reconstructed \mbox{$r^2$-scaled} total intensity image which is obtained by applying a correction for the degree of polarization on the $Q_\phi$ image. All images show a $2\ffarcs0 \times 2\ffarcs0$ field of view on the same linear color scale with equal minimum and maximum value. Orange corresponds to positive values, blue to negative values, and black is the zero-point. The 155~mas diameter coronagraph has been masked out. The contours of the left and center image show the radial distance from the central star to the point of scattering in the disk surface which is used for the \mbox{$r^2$-scaling}. The contours of the right image show the local scattering angles that are used to calculate the phase function and estimate the total intensity.}
\label{fig:image}
\end{figure*}

Here, we will lay out a numerical method to retrieve \mbox{\mbox{$r^2$-scaled}} images and phase functions, taking into account how a flaring disk surface projects onto the image plane. We approximate the height of the scattering surface of the protoplanetary disk, $h_{\rm \tau=1}(r_{\rm mid})$, with a power law function:
\begin{equation}\label{eq:powerlaw}
h_{\rm \tau=1}(r_{\rm mid})=h_0 r_{\rm mid}^\chi,
\end{equation}
where $h_0$ is a normalization constant for the disk height, $r_{\rm mid}$ the disk radius in the midplane, and $\chi$ the power law exponent that determines if the $\tau=1$ surface has a flat ($\chi=1$) or flaring ($\chi > 1$) shape. We assume that the disk is axisymmetric and rotate the radial $\tau=1$ profile around the $z$-axis from which we obtain a grid of coordinates in the disk surface that is visible in scattered light. Next, we project the disk surface onto the image plane for a given disk inclination, $i$. The image plane, $(x_{\rm im},y_{\rm im})$, is placed along the negative $y$-axis such that a point on the disk surface at midplane radius $r_{\rm mid}$, azimuthal angle $\phi$, and disk height $h_{\rm \tau=1}(r)$ projects to the image plane as (see Fig.~\ref{fig:schematic})
\begin{equation}
\begin{aligned}
x_{\rm im} &= r_{\rm mid} \sin{\phi}, \\
y_{\rm im} &= h_{\rm \tau=1}(r) \sin{i} - r_{\rm mid} \cos{\phi} \cos{i}.
\end{aligned}
\end{equation}
The image is then rotated in order to align the major axis position angle, PA, with the observation:
\begin{equation}
\begin{aligned}
x_{\rm im, rot} &= x_{\rm im} \cos{\Theta} - y_{\rm im} \sin{\Theta}, \\
y_{\rm im, rot} &= x_{\rm im} \sin{\Theta} + y_{\rm im} \cos{\Theta}, \\
\end{aligned}
\end{equation}
where $\Theta={\rm PA}-90\degr$ is the image rotation angle and $(x_{\rm im, rot},y_{\rm im, rot})$ are the new image plane coordinates. Since the projection of the coordinates in the disk surface results in an unevenly spaced sampling of the image plane, we use linear interpolation to resample the image plane onto an evenly spaced (10~mas) grid of points in $x_{\rm im,rot}$ and $y_{\rm im,rot}$ direction, which correspond to right ascension and declination respectively.

Now that we know how the disk surface projects onto the detector plane and vice versa, we can map physical quantities to the observed scattered light image. For example, the distance from the star to the point of scattering in the disk surface where $\tau=1$ in radial direction is given by
\begin{equation}\label{eq:scattering_radius}
r = \sqrt{r_{\rm mid}^2+h_{\rm \tau=1}(r_{\rm mid})^2},
\end{equation}
and the scattering angle, $\Psi$, by which stellar photons scatter towards the image plane is defined as
\begin{equation}\label{eq:scattering_angle}
\cos{\left(\pi - \Psi\right)} = \sin{\left(\frac{\pi}{2}+\gamma\right)} \cos{\left(\pi+\phi\right)} \sin{i} + \cos{\left(\frac{\pi}{2}+\gamma\right)}\cos{i},
\end{equation}
with the disk opening angle given by
\begin{equation}
\gamma=\arctan{\left( \frac{h_{\rm \tau=1}(r_{\rm mid})}{r_{\rm mid}} \right)}.
\end{equation}
For an observed scattered light image, we use the right ascension and declination pixel coordinates to interpolate the model image plane. This provides an estimate of the scattering radius and scattering angle in each pixel which can be used to calculate \mbox{$r^2$-scaled} images and dust phase functions.

\section{A new view on the HD~100546 disk surface}\label{sec:HD100546}

As application for the scattered light mapping method, we will use archival polarimetric imaging observations of the protoplanetary disk around HD~100546. The data includes \mbox{SPHERE} \citep{beuzit2008} PDI observations with the Very Large Telescope (VLT) in $R'$-band ($\lambda_{\rm c}= 0.63$ \SI{}{\micro\meter}) that were obtained in 2015 with the optical sub-instrument \mbox{ZIMPOL} \citep{thalmann2008} and presented by \citet{garufi2016}. Furthermore, it includes \mbox{VLT/NACO} \citep{lenzen2003} $H$-band ($\lambda_{\rm c}= 1.66$ \SI{}{\micro\meter}) and $K_{\rm s}$-band ($\lambda_{\rm c}= 2.18$ \SI{}{\micro\meter}) PDI observations from \citet{avenhaus2014} that were obtained in 2013.

HD~100546 is a Herbig~Be star at a distance of $97 \pm 4$ pc \citep{vanleeuwen2007} which is surrounded by a protoplanetary disk \citep{pantin2000}. The disk inclination and major axis position angle have been measured in several studies, for example, $i=41\ffdeg94\pm0\ffdeg03$ and ${\rm PA}=145\ffdeg14\pm0\ffdeg04$ \citep{pineda2014}, $i=44\degr\pm3\degr$ and ${\rm PA}=146\degr\pm4\degr$ \citep{walsh2014}, $i=42\degr\pm5\degr$ and ${\rm PA}=145\degr\pm5\degr$ \citep{ardila2007}. The near side of the disk is most likely along the southwest minor axis (${\rm PA} = 235\degr$) and the far side along the northeast minor axis (${\rm PA } = 55\degr$). This is inferred from the preceding and receding CO lines \citep{pineda2014} and from assuming that the observed spiral arms are trailing \citep{ardila2007,boccaletti2013,avenhaus2014}. The 15~au cavity edge in the \mbox{SPHERE} $R'$-band scattered light image shows no indication of an offset with respect to the central star \citep{garufi2016}.

\subsection{Stellar irradiation corrected images}\label{sec:r2_images}

\begin{figure*}
\centering
\resizebox{\hsize}{!}{\includegraphics{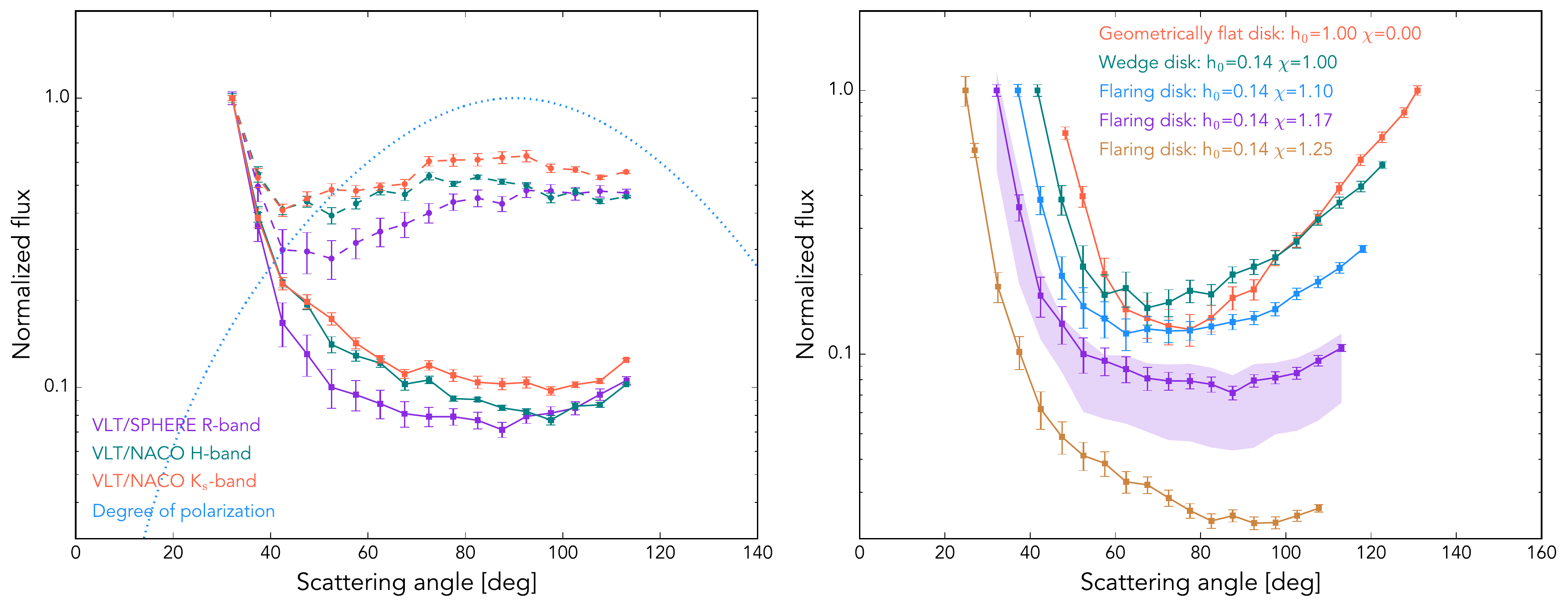}}
\caption{\textbf{Left:} Observed polarized intensity phase functions (dashed lines) and reconstructed total intensity phase functions (solid lines) which are calculated by assuming a bell-shaped degree of polarization (dotted line). The polarized intensity phase functions have been obtained from the $Q_\phi$ scattered light images in $R'$-, $H$-, and $K_{\rm s}$-band. Each phase function has been normalized to its peak value. The error bars show three standard errors of the mean obtained from the corresponding $U_\phi$ images. \textbf{Right:} Total intensity phase functions of the VLT/SPHERE $R'$-band observations for different geometries of the radial $\tau=1$ disk surface (see Eq.~\ref{eq:powerlaw}). Throughout this work, we use $h_0=0.14$ and $\chi=1.17$ (purple data points) which is obtained from the radiative transfer model of \citet{mulders2013}. The purple shaded region shows the effect of the inclination ($40\degr$--\,$45\degr$) on the derived phase function for the $\tau=1$ geometry with $h_0=0.14$ and $\chi=1.17$.}
\label{fig:phasefunction}
\end{figure*}

PDI observations measure the linear polarization components of the Stokes vector, $Q$ and $U$, which are often converted into their azimuthal counterparts, $Q_\phi$ and $U_\phi$. $Q_\phi$ contains all polarized flux in the single scattering limit with only positive pixel values in case of a noise free disk detection with positively polarizing dust grains. Scattered light images are often scaled with the square of the distance from each pixel to the star in order to correct for the dilution of stellar radiation field. This provides a better representation of the spatial distribution of the dust grains in the disk surface and enhances faint structures at large disk radii.

The \mbox{$r^2$-scaled} image is usually obtained by calculating the deprojected distance from each pixel to the star in a geometrically flat disk, for a given inclination and position angle, while the vertical extent of the disk is neglected \citep[e.g.,][]{avenhaus2014,garufi2016}. In that case, the distance calculation is symmetric with respect to the major axis and pixels that are on opposite sides of the major axis are being scaled by the same amount. This introduces large errors for inclined and flaring disks because the distances on the near side will be underestimated while the distances on the far side will be overestimated. The left image in Fig.~\ref{fig:image} shows the \mbox{$r^2$-scaled} $R'$-band $Q_\phi$ image with only a correction applied for the inclination of the HD~100546 disk. In that case, the far side appears brighter in the \mbox{$r^2$-scaled} image compared to the near side of the disk \citep{avenhaus2014,garufi2016}.

For a more realistic $r^2$-scaling, we use the \mbox{MCMax} \citep{min2009} radiative transfer model from \cite{mulders2013} to estimate the height of the $\tau=1$ surface of the HD~100546 disk. The radiative transfer model was constructed to fit \emph{Hubble Space Telescope} (HST) scattered light images in the optical and near-infrared, as well as the spectral energy distribution (SED). We retrieved the radial $\tau=1$ height from the model in $R'$-, $H$-, and $K_{\rm s}$-band and performed a least squares fit of Eq.~\ref{eq:powerlaw} from which we obtained $h_0 = 0.14$ and $\chi=1.17$ for the three filters. The $\tau=1$ height is very similar in $R'$-, $H$-, and $K_{\rm s}$-band, with differences in $h_0$ and $\chi$ only after the second decimal. We note that the flaring of the $\tau=1$ surface ($\chi=1.17$) is smaller than the flaring of the pressure scale height, $H_{\rm p}=0.04\,{\rm au}\,(r/{\rm au})^{1.3}$ \citep{mulders2013}, because the height of the $\tau=1$ surface is determined by the combined effect of the parametrized pressure scale height and surface density.

The center image in Fig.~\ref{fig:image} shows the $Q_\phi$ image with a \mbox{$r^2$-scaling} that is calculated with the power law profile that was obtained from the radiative transfer model. Each pixel of the original $Q_\phi$ image has been multiplied with the squared distance from the star to the point of scattering in the disk surface as given by Eq.~\ref{eq:scattering_radius}. The resulting \mbox{$r^2$-scaled} image appears significantly different from the one constructed with the geometrically flat assumption with the near side (southwest) of the disk clearly being brighter than the far side (northeast). The reason for this is apparent from the scattering radius contours on the left and center image of Fig.~\ref{fig:image}. The contours are symmetric with respect to the disk major axis for the geometrically flat case. In reality, the scattering distance is projected asymmetrically with respect to the disk major axis because of the flaring geometry of the disk surface.

\subsection{Dust phase functions}\label{sec:phase_function}

The scattering phase function of dust grains depends on their size, shape, composition, and internal structure among others \citep[e.g.,][]{bohren1983,munoz2006,min2012,min2016}. To gain insight into the scattering properties of the dust grains residing in the HD~100546 disk surface, we obtained the phase function from the unscaled $Q_\phi$ polarized intensity images. In Sect.~\ref{sec:r2_images}, we used the scattered light mapping method to determine the scattering radius and scattering angle associated with each image pixel (see contours on the center and right image of Fig.~\ref{fig:image}, respectively). This allows us to construct a phase function from all pixel values within a certain distance range from the star.

Figure \ref{fig:phasefunction} shows the derived phase functions for the $R'$-, $H$-, and $K_{\rm s}$-band PDI observations. We have used all the pixels that correspond to disk radii between 80 and 100~au, corrected each pixel value for the effect of stellar irradiation, and determined the scattering angle for each pixel with Eq.~\ref{eq:scattering_angle}. The pixel values are binned in 36 linearly spaced scattering angle bins between $0\degr$ and $180\degr$. The polarized intensity phase functions (dashed lines in Fig.~\ref{fig:phasefunction}) are constructed from the mean scattering angle and the mean $r^2$-scaled $Q_\phi$ flux in each bin. The uncertainties are calculated as $3\sigma/\sqrt{N}$, with $\sigma$ the standard deviation and $N$ the number of pixels, from the corresponding $r^2$-scaled $U_\phi$ images.

The polarized intensity phase function is controlled by the combined effect of the total intensity phase function and the degree of polarization. In the single scattering limit ($Q_\phi$ is dominated by single scattered light), the degree of polarization is given by the single scattering polarization which in case of silicate dust grains can often be approximated by a bell-shaped curve, both for small and large grains, but also spherical or aggregate in structure \citep[e.g][]{mishchenko2000,mishchenko2002,volten2001,murakawa2010,min2005,min2016}. Therefore, we use the Rayleigh single scattering polarization,
\begin{equation}\label{eq:rayleigh}
P = -\frac{\cos^2{\Psi}-1}{\cos^2{\Psi}+1},
\end{equation}
to estimate the total intensity phase functions (solid lines in Fig.~\ref{fig:phasefunction}) by dividing the polarized intensity phase functions with the single scattering polarization. We note that normalizing the bell-shaped polarization curve to a lower peak value will not affect the phase function because it is shown in normalized flux units. The phase functions show an isotropic part for scattering angles in the range of $50\degr$--$100\degr$ and a steeply rising part below $50\degr$ \citep[see also][]{ardila2007}. The lower and upper limit scattering angles that are probed in the HD~100546 disk surface are approximately $32\degr$ and $113\degr$, respectively. The overall shape of the derived phase functions is similar in $R'$-, $H$-, and $K_{\rm s}$-band, although the steepness of the forward scattering slope increases towards shorter wavelengths.

Since we have estimated the degree of polarization in each pixel (neglecting multiple scattering effects), we can reconstruct a total intensity image of the disk by dividing each $Q_\phi$ pixel value with the degree of polarization. We have normalized the bell-shaped polarization curve to a peak value of 0.5 which is typical for compact aggregate dust grains \citep{min2016}. The single scattering polarization of fluffy aggregates will also be bell-shaped, although the peak value will be higher than for compact aggregates \citep{tazaki2016}. The \mbox{$r^2$-scaled} total intensity image with an contour overlay of the scattering angles is shown on the right of Fig.~\ref{fig:image}. The image is shown with the same color scaling as the \mbox{$r^2$-scaled} $Q_\phi$ images and shows that the near side of the disk is very bright in total intensity due to the forward scattering nature of the dust grains. The \mbox{$r^2$-scaled} scattered light decreases continuously in azimuthal direction away from the near side minor axis and shows a slight minimum around scattering angles of $100\degr$. Towards the far side minor axis, there is again a slight increase in brightness, possibly by the beginning of the backward scattering peak of the phase function (see Sect.~\ref{sec:discussion_aggregates}) and/or an intrinsically enhanced surface brightness from an asymmetry in the disk structure \citep{garufi2016}.

\section{Discussion}\label{sec:discussion}

\subsection{Interpretation of scattered light images}

Application of the scattered light mapping method on the PDI observations of HD~100546 has shown that the flaring shape of the radial $\tau=1$ surface significantly affects the \mbox{$r^2$-scaled} images and phase functions. More generally, using a geometrically flat disk for the \mbox{$r^2$-scaling} might lead to a misinterpretation in case the disk surface is inclined and flaring. The calculated \mbox{$r^2$-scaled} $Q_\phi$ image (see center image of Fig.~\ref{fig:image}) and the derived phase functions (see left plot of Fig.~\ref{fig:phasefunction}) differ significantly from earlier results in the literature. Specifically, previous studies seemed to show \mbox{$r^2$-scaled} $Q_\phi$ images that are brighter on the far side of the disk than the near side which has been explained by a backward peak of the phase function whereas the beginning of the forward peak of the phase function would explain the small amount of scattered light flux from the near side \citep{quanz2011,avenhaus2014,garufi2016}. Our results show the opposite: the near side of the disk is brighter in the \mbox{$r^2$-scaled} $Q_\phi$ image than the far side.

With the scattered light mapping, we find that part of the forward scattering peak of the phase function is probed on the near side of the disk and possibly the beginning of the backward scattering peak is visible on the far side of the disk. The two dark regions in the \mbox{$r^2$-scaled} $Q_\phi$ image that are aligned with the major axis (see Fig.~\ref{fig:image}) are likely the result of the combined effect of the total intensity phase function and the degree of polarization. More specifically, the total intensity phase function peaks in forward direction (the near side of the disk) while the degree of polarization peaks at a $90\degr$ scattering angle (near the major axis). This results in a minimum in polarized intensity around scattering angles of $50\degr$. This is clearly seen from the minimum in the polarized intensity phase functions in the left plot of Fig.~\ref{fig:phasefunction}.

An alternative explanation for the low surface brightness regions in the \mbox{$r^2$-scaled} $Q_\phi$ image is a scenario in which a warped inner disk is casting shadows on the outer disk \citep{avenhaus2014,garufi2016}. This seems unlikely for two reasons. Firstly, a shadow that would be cast by a misaligned inner disk is expected to be visible on the outer disk rim as well which is not the case here \citep{marino2015,stolker2016}. Secondly, the boundary of the low surface brightness regions is on one side approximately aligned with the major axis of the outer disk which would require the inner disk to have the same position angle as the outer disk which would be coincidental. Future observations will establish if the HD~100546 polarized surface brightness pattern is also visible in other, similarly inclined and flaring disks.

\subsection{Evidence for aggregate dust grains}\label{sec:discussion_aggregates}

The derived total intensity phase functions of the HD~100546 disk are presented in the left plot of Fig.~\ref{fig:phasefunction} and show an overall similar shape in $R'$-, $H$- and $K_{\rm s}$-band. The steeply rising forward scattering peak below scattering angles of $50\degr$ and the isotropic phase function at intermediate scattering angles indicate that the dust grains in the disk surface are large ($2\pi a\gtrsim\lambda$, with $a$ the grains radius and $\lambda$ the photon wavelength). Smaller grains have a more moderate forward scattering peak or scatter approximately isotropic at all scattering angles \citep[e.g.,][]{min2016}. The phase functions show an increase in the steepness of the forward scattering peak towards smaller wavelengths which is expected because the dust grains will become larger compared to the photon wavelength.

Large compact grains settle efficiently towards the midplane \citep{dubrulle1995}, therefore it is expected that the dust grains in the disk surface of HD~100546 will have an aggregate structure which provides them with aerodynamic support against vertical settling. This is consistent with the results from \citet{mulders2013} who have shown that the red color, low albedo, and small brightness asymmetry of total intensity scattered light observations with HST are an indication for large aggregate dust grains in the disk surface of HD~100546.

The scattered light observations of HD~100546 probe only part of the dust phase function and in particular large scattering angles are not observed. The reconstructed total intensity phase functions (see left plot of Fig.~\ref{fig:phasefunction}) show the beginning of the backward scattering peak as expected for large aggregate dust grains \citep{min2016,tazaki2016}. However, the presence and strength of the observed backward peak depends on the assumed geometry of the disk surface. The right plot in Fig.~\ref{fig:phasefunction} shows the derived phase function from the $R'$-band image for different power law profiles of the disk height which shows a large difference between a geometrically flat and a flaring surface.

For this analysis, we have assumed that the observed flux is dominated by single scattered light. A more detailed radiative transfer study is required to investigated the effect of multiple scattering on the derived phase function. Also, we have neglected the effect of dust composition, which especially has an effect on the single scattering polarization \citep[e.g.,][]{munoz2006}. However, the dust composition of the disk is likely dominated by silicate materials which all have comparable refractive indices and single scattering polarization curves that are approximately bell-shaped \citep[e.g.,][]{volten2001}. Nevertheless, the assumption of a bell-shaped single scattering polarization may have introduced a bias in the analysis.

\section{Conclusions}\label{sec:conclusions}

We present a numerical method, scattered light mapping, for the interpretation of scattered light images of protoplanetary disks. The method considers how a flaring disk surface projects on the image plane of the observer. In this way, the scattering radii and scattering angles can be mapped to the pixels of a scattered light image such that stellar irradiation corrected images and phase functions can be retrieved. The main conclusions are:

\begin{itemize}
  \item Taking into account the projection effect of a inclined and flaring disk surface on the image plane can have a significant effect on the calculation of stellar irradiation corrected images and dust phase functions. An estimate of the height of the $\tau=1$ surface is required which may be obtained from a radiative transfer model that reproduces, at minimum, the infrared excess of the SED, or from fitting ellipses to concentric structures, such as rings and gaps, in a scattered light image of an inclined disk \citep{ginski2016}.
  \item Application of the mapping method on polarized scattered light images of the protoplanetary disk around HD~100546 revealed that the near side is brighter than the far side both in polarized intensity and total intensity, as opposed to earlier work with results inferred from a geometrically flat disk geometry.
  \item The derived dust phase functions from the $R'$-, $H$, and $K_{\rm s}$-band $Q_\phi$~images of HD~100546 indicate that large ($2\pi a\gtrsim\lambda$) dust grains dominate the scattering opacity in the disk surface. Large dust grains in the disk surface are expected to have an aggregate structure which prevents them from settling efficiently towards the midplane.
\end{itemize}

\begin{acknowledgements}
We are thankful to the anonymous referee for providing valuable comments. H.A. acknowledges support from the Millennium Science Initiative (Chilean Ministry of Economy), through grant "Nucleus RC130007" and from FONDECYT grant 3150643. SPHERE is an instrument designed and built by a consortium consisting of IPAG (Grenoble, France), MPIA (Heidelberg, Germany), LAM (Marseille, France), LESIA (Paris, France), Laboratoire Lagrange (Nice, France), INAF Osservatorio di Padova (Italy), Observatoire de Geneve (Switzerland), ETH Zurich (Switzerland), NOVA (Netherlands), ONERA (France) and ASTRON (Netherlands) in collaboration with ESO. SPHERE was funded by ESO, with additional contributions from CNRS (France), MPIA (Germany), INAF (Italy), FINES (Switzerland) and NOVA (Netherlands). SPHERE also received funding from the European Commission Sixth and Seventh Framework Programmes as part of the Optical Infrared Coordination Network for Astronomy (OPTICON) under grant number RII3-Ct-2004-001566 for FP6 (2004-2008), grant number 226604 for FP7 (2009-2012) and grant number 312430 for FP7 (2013-2016). This research made use of Astropy, a community-developed core Python package for Astronomy \citep{astropy2013}.
\end{acknowledgements}

\bibliographystyle{aa}
\bibliography{references}

\end{document}